# Using Silent Writes in Low-Power Traffic-Aware ECC

Mostafa Kishani[1], Amirali Baniasadi[2], Hossein Pedram[1]

[1]Department of Computer Engineering and Information Technology,
Amirkabir University of Technology, Iran
[2]Electrical & Computer Engineering, University of Victoria, Canada
mkishani@aut.ac.ir, amirali@ece.uvic.ca, pedram@aut.ac.ir

**Abstract.** Using Error Detection Code (EDC) and Error Correction Code (ECC) is a noteworthy way to increase cache memories robustness against soft errors. EDC enables detecting errors in cache memory while ECC is used to correct erroneous cache blocks. ECCs are often costly as they impose considerable area and energy overhead on cache memory. Reducing this overhead has been the subject of many studies. In particular, a previous study has suggested mapping ECC to the main memory at the expense of high cache traffic and energy. A major source of this excessive traffic and energy is the high frequency of cache writes. In this work, we show that a significant portion of cache writes are silent, i.e., they write the same data already existing. We build on this observation and introduce Traffic-aware ECC (or simply TCC). TCC detects silent writes by an efficient mechanism. Once such writes are detected updating their ECC is avoided effectively reducing L2 cache traffic and access frequency. Using our solution, we reduce L2 cache access frequency by 8% while maintaining performance. We reduce L2 cache dynamic and overall cache energy by up to 32% and 8%, respectively. Furthermore, TCC reduces L2 cache miss rate by 3%.

## 1 Introduction

Memory reliability impacts the overall processor reliability significantly. This is due to the fact that an error in memory can easily propagate to other system elements and influence both data results and control flow. Soft errors, i.e. errors which are transient and change one or more bit values transiently (unlike permanent errors which are steady forever), are the main cause of errors in memory [12, 13]. Cache memories deserve special attention as they are the closest memory layer to the CPU, effectively having the most data exchange. Consequently, any error occurring in the cache memories is likely to propagate into the CPU structures. Moreover, cache memories use SRAM cells and are therefore very susceptible to errors (unlike main memory which uses DRAM cells [14]). Soft error probability has remained the same for SRAM cells [15, 16, 17, 18]. Meantime, cache sizes continue to grow increasing the overall likelihood of soft error in cache memories.

In order to protect the system against such errors, error checking and correcting schemes are used to prevent errors from propagating to other parts of system. Exploiting *Errors Detection Codes* (EDCs) and *Error Correction Codes* (ECCs) is one way to achieve error protection. EDCs are often low cost codes such as parity bits capable of detecting single or multiple bit flips in a cache block. Conventionally, EDCs are used to check data integrity on each cache read operation. Compared to EDCs, ECCs (e.g., hamming code) come with higher complexity (and hence overhead) as they require error correction capabilities. Previous work has suggested many solutions to reduce ECC overhead [1, 2, 3, 4, 5, 6]. In particular, a previous study [1], has suggested mapping ECC to the main memory as regular data, referred to also as MMECC. MMECC reduces area and leakage energy overhead at the cost of low performance degradation. On the negative side, MMECC increases both cache traffic and dynamic energy as it increases the frequency of cache write operations (as it writes the ECC value using an extra cache access).

In this paper, we extend previous work and use the observation that a considerable portion of cache writes write the data already existing. We refer to this group of writes as *silent writes*. Conventionally, cache blocks written by a silent write are treated as dirty blocks. This imposes additional unnecessary overhead calculating, updating and rewriting ECC values. As we show in this work detecting and avoiding these redundant computations can save energy and traffic in cache memory. To this end, we introduce Low-Power Traffic-Aware ECC (TCC). By detecting and skipping conventional steps for silent writes, TCC reduces both cache traffic and energy consumption. TCC relies on an efficient and effective mechanism to detect silent writes. This is done by comparing low cost signatures associated with each block to find out if a write is in fact silent. To minimize signature cost, we use the already existing parity code as block signature and show that 98% of the non-silent writes can be detected by using parity. In particular we make the following contributions:

- We show that a significant portion of cache writes to the L2 cache are silent writes. We show that in the case of silent write, writing the cache block and updating the associated ECC can be avoided.
- We introduce an efficient mechanism to detect silent writes by using the already existing parity code as block signature. We show that 98% of non-silent writes can be detected by using parity code as signature.
- By skipping ECC calculation and update for silent writes, we reduce cache access frequency (max: 32%), dynamic energy (max: 32%) and miss rate (max: 3%).

The rest of this paper is organized as follows: Section 2 explains related work. Section 3 describes background information including decoupled EDC and ECC. We review our motivating observations in Section 4. In Section 5 we present TCC in more details. We report methodology and results in Section 6 and 7 respectively. Finally in Section 8, we offer concluding remarks.

## 2 Related Work

Li [5] proposed ECC power gating for clean lines to reduce leakage power. Sadler and Sorin [4] used punctured error codes instead of conventional hamming codes for error detection and correction. Punctured code can be separated to EDC and ECC parts. In order to save cache read latency, they proposed decoupling EDC from ECC and suggested using a dedicated cache called Punctured ECC Recovery Cache (PERC) to hold the ECC part of the code. Meantime, EDC is held in the data cache. In order to reduce area and power, Kim [6] suggested using up to one dirty line per cache set, storing ECC only for this single dirty line.

Yoon and Erez [1] suggested mapping ECC to the memory system (MMECC), instead of storing it at the end of the cache line. MMECC saves cache space by writing ECC in the memory space. This does not impact memory traffic significantly as ECC is only updated when necessary. Meantime, MMECE comes with an additional space overhead in the last cache level. Furthermore, MMECC increases the traffic of the last level cache as a result of regular ECC updates. These consequences could potentially have a negative impact on both power and performance. In this work we show that taking application behavior into account can reduce the cache traffic overhead associated with MMECC. Lepak and Lipasti [7] introduced the concept of silent stores as store instructions writing the same value already stored. Zhang [2] proposed In-Cache Replication (ICR) for L1 data caches. ICR uses existing cache space to hold replicas of cache blocks. Kim and Somani [3] proposed parity caching, shadow checking and selective checking.

Protecting the most error prone blocks (as [5] and [6] do) increases cache reliability. This level of protection, however, may not be sufficient for highly reliable and critical applications. TCC does not compromise error correction capability as it stores ECC for all dirty cache blocks uniformly while imposing little area, power and bandwidth overhead. We reduce both the area and the leakage energy overhead associated with ECC by storing the required information in the memory (similar to [1]) rather than a dedicated cache (as [4], [5] and [6] do). Meantime, we improve MMECC [1] by reducing the associated traffic overhead by limiting the L2 cache writes to the writes that are not silent.

## 3 Decoupled EDC and ECC Background

On a cache read operation, EDC is read and calculated to check block integrity to assure correctness. Cache read operations are on the critical path and occur frequently. Therefore, implementing EDC operations efficiently would enhance both the overall performance and energy. While hamming code has proven to be a viable solution to detect and correct errors in caches, it suffers from significant overhead. Consequently, designers have preferred fast and low demanding error detection mechanisms like parity. It is important to note that ECC is not needed in write-through caches, as there is always an intact copy of the data block available in the upper memory level. In the case of a write-back cache, however, the upper level only includes a copy of the clean data, making dirty blocks that are not yet evicted, vulnerable to errors. To protect write-back caches against errors we use ECC if EDC shows an error. Since correction needs a more powerful code, write-back caches exploit two different codes, a low-power and low-latency code (like parity) for error detection, and a powerful ECC for correction (like hamming). EDC and ECC are calculated and stored on a dirty write-back into the cache. When EDC detects an error, if the cache line is clean the correct block is read from the higher memory level. Otherwise, ECC is used to produce the correct cache block. One way to store EDC and ECC is to place them at the end of the cache line. This, while easy for EDC, is not affordable for ECC as ECC comes with significant memory overhead. For example, a cache using a single error correction hamming code requires eight bytes of overhead for a 64-byte block. The associated increase in the cache line size increases the cache area and energy (both dynamic and leakage).

## 4 Motivation

We are motivated by the fact that a significant portion of consecutive writes on the L2 and L1 caches are silent, i.e., they rewrite the same block written previously, consuming energy and bandwidth without contributing to performance. As presented, on average, silent writes account for 37% of L2 cache writes. In **Fig. 1** we report the share of silent writes (see Section 6 for methodology). In addition, and in a system using MMECC, silent writes result in calculating and rewriting already stored parity and hamming bits hence furthermore wasting activity. Identifying and avoiding this redundant activity improves energy efficiency in two ways.

- Since previous and new ECCs are equal, calculating the hamming code can be eliminated for silent writes. Note that the XOR operations required to calculate the hamming code are four times more than the XOR operations needed for block comparison.
- In methods like Memory Mapped ECC [1], ECC should be written to memory or data cache on each cache write operation. This could be avoided for silent writes, reducing cache access activity.

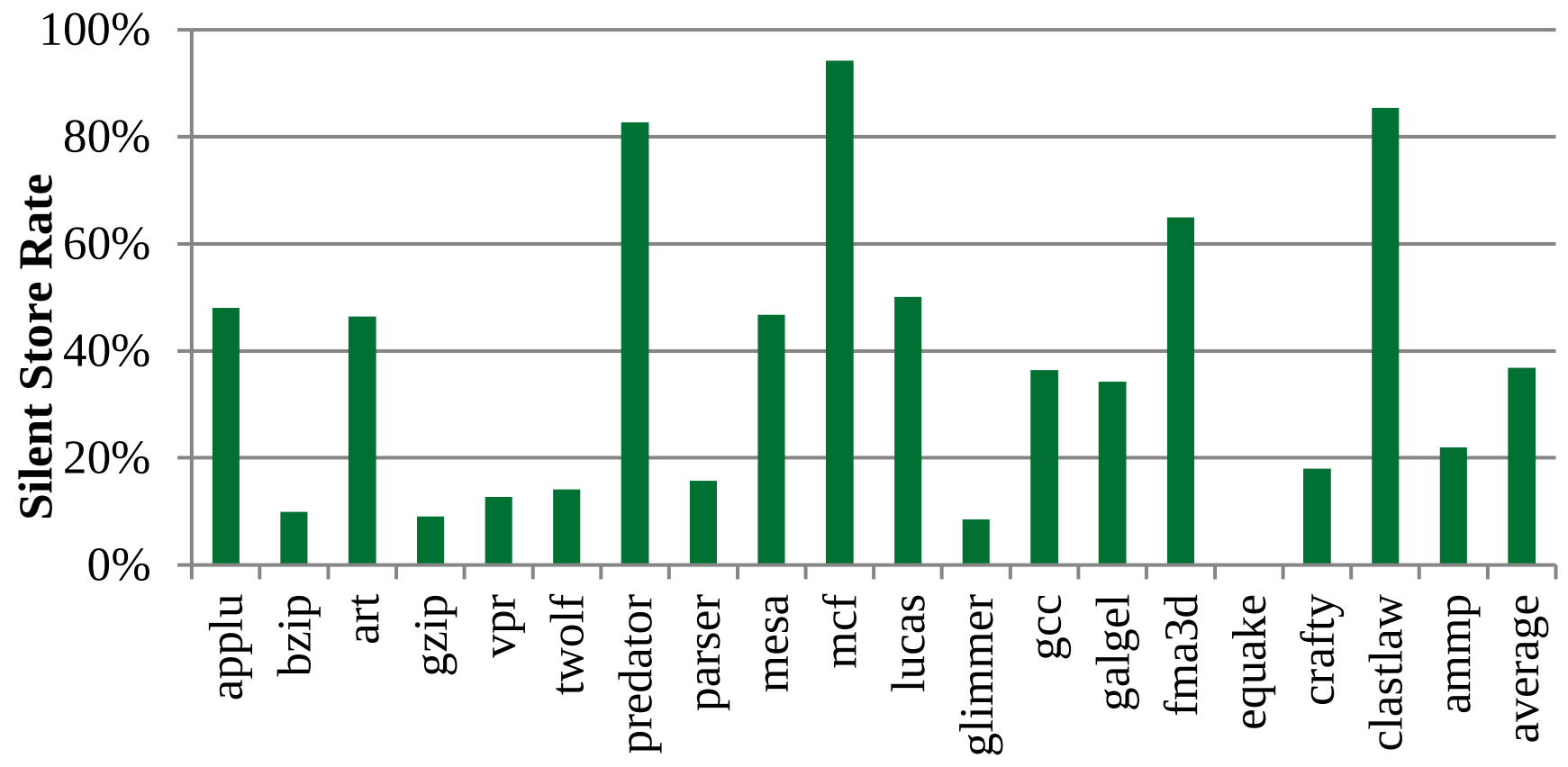

**Fig. 1.** Silent write frequency for L2 cache.

## 5 TCC

MMECC requires frequent cache write operations as it writes the ECC associated with each cache block either to the memory or to the cache upon each data cache write. This is unnecessary as cache writes rewrite the same data frequently as presented in Section 4. We take advantage of this observation and suggest Traffic-Aware ECC, or simply TCC. TCC uses application behavior to reduce cache/memory traffic and eliminate a significant share of the hamming calculations and cache updates performed by conventional ECC solutions.

TCC is applied to a memory system already decoupling error detection from error correction [4] and utilizing memory mapping [1] (requiring treating ECC as a regular data). TCC uses interleaved parity as EDC. ECC is computed and written upon a cache write-back from L1 to L2. Parity comes with one byte overhead and is stored at the end of each cache line in TCC. Hamming, on the other hand, has an eight byte overhead per 64-byte size cache line and is stored in the memory.

TCC allocates an address location to the ECCs and maps the ECC of each cache block to a memory address. To accommodate ECC, and similar to MMECC, TCC stores 8 bytes of ECC as the block-ECC for the 64-byte cache block used in this study. We group eight block-ECCs to form a memory block of 64 bytes. We refer to the cache blocks associated with each of the block-ECCs stored in one memory block as adjacent blocks. In an 8-way set associative cache, cache lines in the eight consecutive cache sets with the same way number form the adjacent blocks. For instance, cache lines in the way 0 of set0 to set7, are adjacent blocks.

In the case of cache writes, if the ECC line is cached, the new ECC should be stored using an extra cache write. Otherwise, a cache block is assigned to the associated ECC. In case one or more adjacent blocks are dirty, their block-ECC has already been saved in the memory (as we just save ECC for dirty cache blocks). Here we read the already stored ECC block from memory.

## 5.1 Block Comparison

Detecting silent writes needs comparing each ready-to-write block to the old block. This requires reading the old block dissipating power comparable to that of a cache write. For TCC to achieve its goal, it is important to perform this comparison efficiently as suggested in the next section.

**Detecting Silent Writes.** To reduce the overhead associated with silent write detection, we exploit low cost small signatures (explained in the next section) associated with each block rather than comparing the entire block addresses. To this end we use a small dedicated cache, referred to as the signature cache. The number of signature cache lines is the same as the L1 cache and the line size is one byte. When a data block is written from L2 to L1, its signature is calculated and saved in the signature cache.

At the time of a write back from L1 to L2, the signature of the ready-to-write block is calculated and compared to the old signature saved in the signature cache. If different, the write is not a silent write (i.e. the new and old data blocks are not the same), hence the ready-to-write block should be written to the L2 cache, and the ECC should be computed and rewritten. If the signatures are equal, there is still a chance that the two blocks are unequal. Therefore, TCC takes an extra step comparing the old block (which is read from the L2 cache) to the ready-to-write block. If equal, the write operation is a silent write and no further action is required, as both blocks and their associated ECCs are equal. Otherwise, the write is not silent. Therefore, the new data block should be written to the L2 cache, and the ECC must be computed and written following the conventional approach. The hardware overhead of comparing the old block to the ready-to-write block is a 64-bit comparator. We take the power overhead associated with this comparator in the energy results presented in Section 7.3. In the next section we present how the signature is calculated.

**Parity as Signature.** The interleaved parity code, which is used for error detection, should be calculated and saved for each cache block. We use the already existing parity bits as signature bits. Using parity as signature has two benefits: a) we no longer need extra effort to calculate signature b) we no longer need extra storage for signature as parity is already stored in the cache block. **Fig. 2** shows the steps TCC takes during L2 cache read and writes.

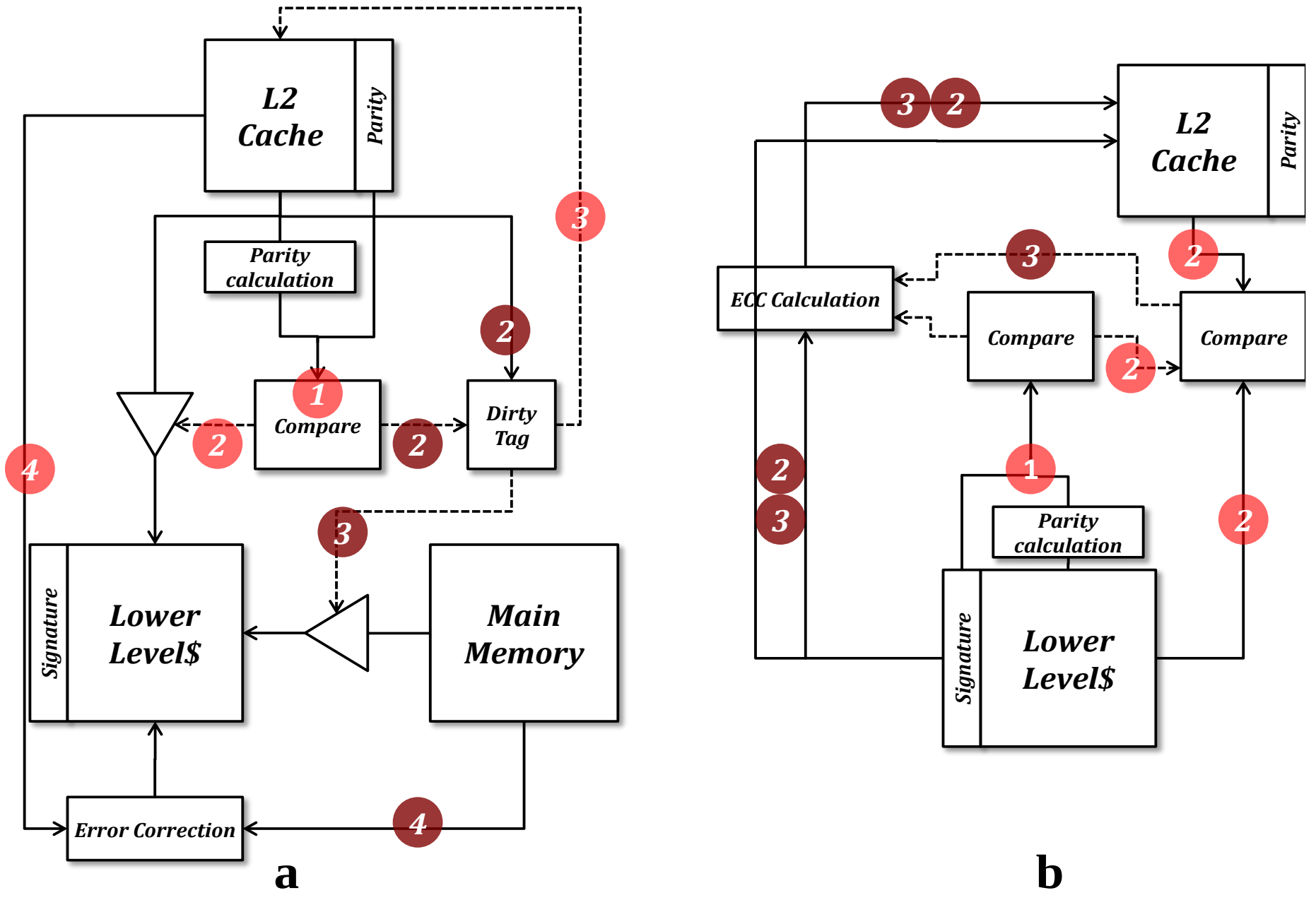

**Fig. 2.** L2 cache read and write mechanisms used in a TCC-enhanced system. (Note that labels with the same number and color happen in parallel. The dashed lines show the control signal) **a:** L2 cache read operation 1) read the data block and the associated parity/signature. Calculate the data block parity independently and compare to the fetched parity. 2) (Lighter label) in the case of a parity match, the block is error-free. Write the block to the lower level cache. 2) (Darker labels) parity mismatch. Check if the block is dirty or not. 3) (Darker label) block is not dirty. Read the correct data from the main memory. 3) (Lighter label) block is dirty. Check the cache for the ECC. 4) (Lighter label) ECC is cached. Read the ECC from cache and correct the data block. 4) (Darker label) ECC is not cached. Read the ECC from the main memory and correct the block. **b:** L2 cache write operation. 1) Read the data block and the associated parity/signature. Calculate its parity and compare to the signature 2) (Darker labels) signature mismatch. So this is a non-silent write, i.e. the ready-to-write block is not equal the previously stored block. Calculate the ECC; write the block and ECC to the cache. 2) (Lighter labels) signature match. Compare current L2 block to the ready-to-write block. If they match, this is a silent write and the write operation is done. 3) (Darker labels) block mismatch. So this is a non-silent write. Calculate the block ECC; write the block and ECC to the cache.

Our study shows that about 98% of the time, unequal data blocks come with unequal associated signatures (i.e., 98% of the time comparing signatures is enough to find out that a write is not silent). Meantime, in 2% of the cases signatures are equal while their associated blocks are different. In this work we propose using parity as signature assuming that the cache architecture holds parity as EDC for each block while using SEC-DED hamming code as ECC. Meantime, our solution can be used for any ECC type which uses parity as EDC. In the event when parity is not used as EDC, other appropriate data representations could be used as signature instead of parity.

## 6 Methodology

We use Simplescalar 2 [8] to evaluate our solutions. We execute 500M representative instructions from SPEC2000 [9] using SimPoint [11]. In order to estimate cache dynamic and leakage energy, we use CACTI 6 [10] tool. Table 1 shows the system configuration used in this study. We assume that both L2 and L1 caches have same block size [20, 21, 22]. Note that we take into account all extra tasks contributing to energy consumption in TCC and MMECC when comparing to a conventional processor. These tasks include:

- Signature read and write
- Signature comparison
- Data block read and write
- Data block comparison

In the conventional method, ECC is stored along the data block, so the length of the data block is more than TCC. Moreover, we assume that parity is stored at the end of the cache block and is read by each cache read operation. We calculated the cost of both data and signature read and update using CACTI tool. Meantime, we use synopsys HSPICE tool [19] for 45nm technology to estimate block and signature comparison cost.

Table 1: CPU Configuration

| **Processor component** | **Value** |
|---|---|
| Integer Functional Unit | 4 ALU, 1 Multiplier/Divider |
| Floating Point Functional Unit | 4 ALU, 1 Multiplier/Divider |
| Instruction Fetch Queue/LSQ/RUU size | 4/32 / 64 Instructions |
| Decode/Issue/Commit Width | 4 / 4 / 4 instructions |
| Memory Latency | First Chunk 512 cycle/Inter Chunk 128 cycle |
| Memory System Ports (to CPU) | 2 |
| Branch Predictor | Comb: 1024 meta size,<br>Bimodal: 2048,<br>2level: 8 bits history and 1024 arra<br>BTB: 512, 4-way |
| Mis-prediction Latency | 3 cycle |
| Level 1 Instruction Cache | 32KB / 3cycles access latency / 4 ways / 64 bytes per block / LRU |
| Level 1 Data Cache | 64KB/3cycles access latency / 4 ways / 64 bytes per block / LRU |
| Level 2 Instruction Cache | 256KB / 12cycles access latency / 4 ways / 64 bytes per block / LRU |
| Level 2 Data Cache | 1MB/12cycles access latency / 8 ways / 64 bytes per block / LRU |

# 7 Results

In this section we report experimental results. We report performance in Section 7.1. We present TCC impact on cache access frequency in Section 7.2. In Section 7.3 we report energy consumption. Finally, we present how TCC impacts cache miss rate in 7.4. To provide better understanding we compare TCC to MMECC [1] and a conventional system.

## 7.1 Performance

In **Fig. 3** we report performance for TCC and MMECC compared to a conventional system storing ECC entirely in a dedicated cache array space. As presented in **Fig. 3**, TCC shows competitive performance compared to MMECC. While both methods show slight performance loss compared to the conventional system, TCC seems to better maintain performance. On average, TCC and MMECC show 0.07% and 0.06% performance loss compared to the conventional system. This performance loss has two reasons. First, unlike the conventional system that stores both ECC and the cache block in the same cache access, MMECC and TCC require an additional cache access to store ECC. Both MMECC and TCC store the data, compute the ECC and then store the ECC in the cache space. This results in extra traffic which can impact performance. Second, we assume that MMECC and TCC use cache space to store ECC while the conventional system stores ECC in a separate array. Consequently, TCC and MMECC increase cache miss rate slightly (see Section 7.4) which has a negative impact on performance.

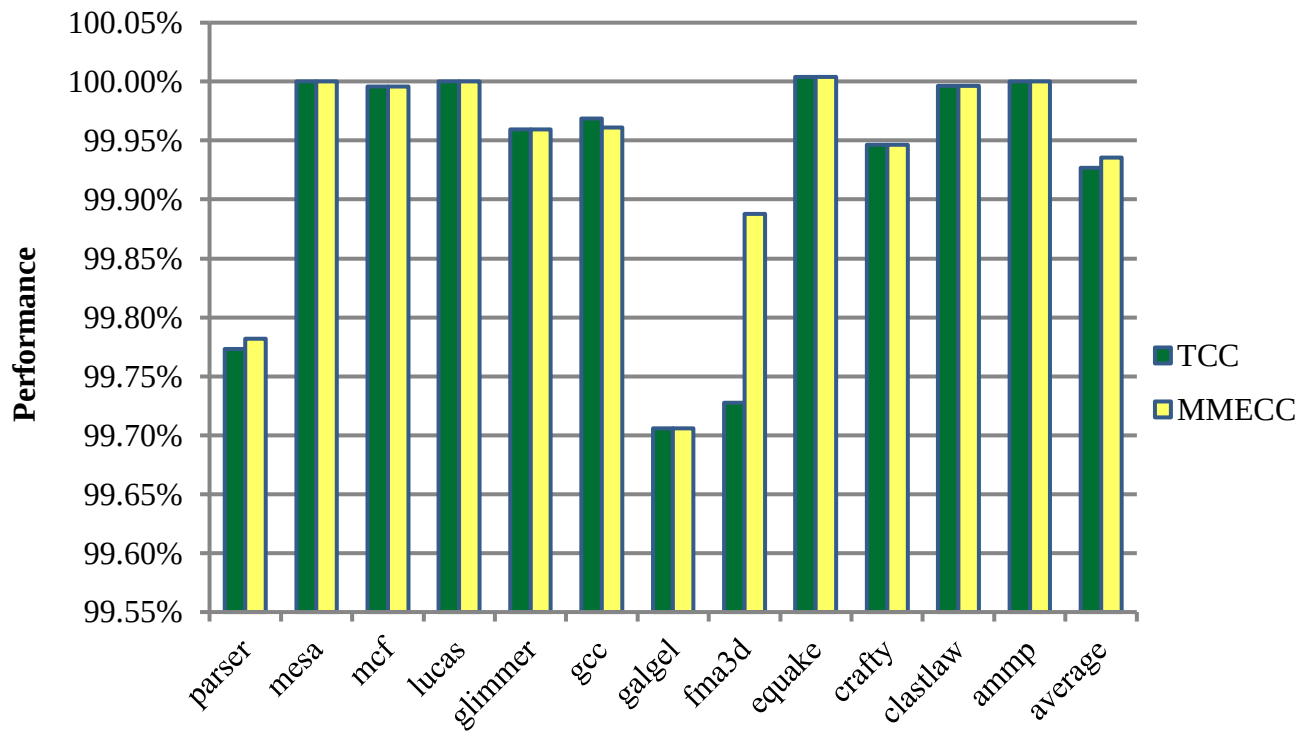

**Fig. 3.** Relative Performance

## 7.2 Cache Access

In **Fig. 4** we report the number of L2 data cache accesses for TCC and MMECC compared to the conventional system. On average, TCC and MMECC show 21% and 32% cache access frequency increase compared to the conventional system respectively. As TCC updates ECC by an extra L2 cache access (except when the new block is equal to previous one), its L2 access is higher than conventional system. TCC shows less L2 access frequency compared to MMECC as we avoid writing the data block and the associated ECC in the case of a silent write. On average, TCC shows 8% (up to 32%) L2 access reduction compared to MMECC. The amount of access frequency difference between TCC and MMECC depends on how often silent writes occur. For example in equake, bzip and glimmer, TCC and MMECC accesses are almost equal. This is consistent with **Fig. 1** where these three benchmarks show a low rate of silent writes. On the other hand, TCC accesses in clastlaw and mcf is much less than MMECC as silent write frequency is high for these two benchmarks.

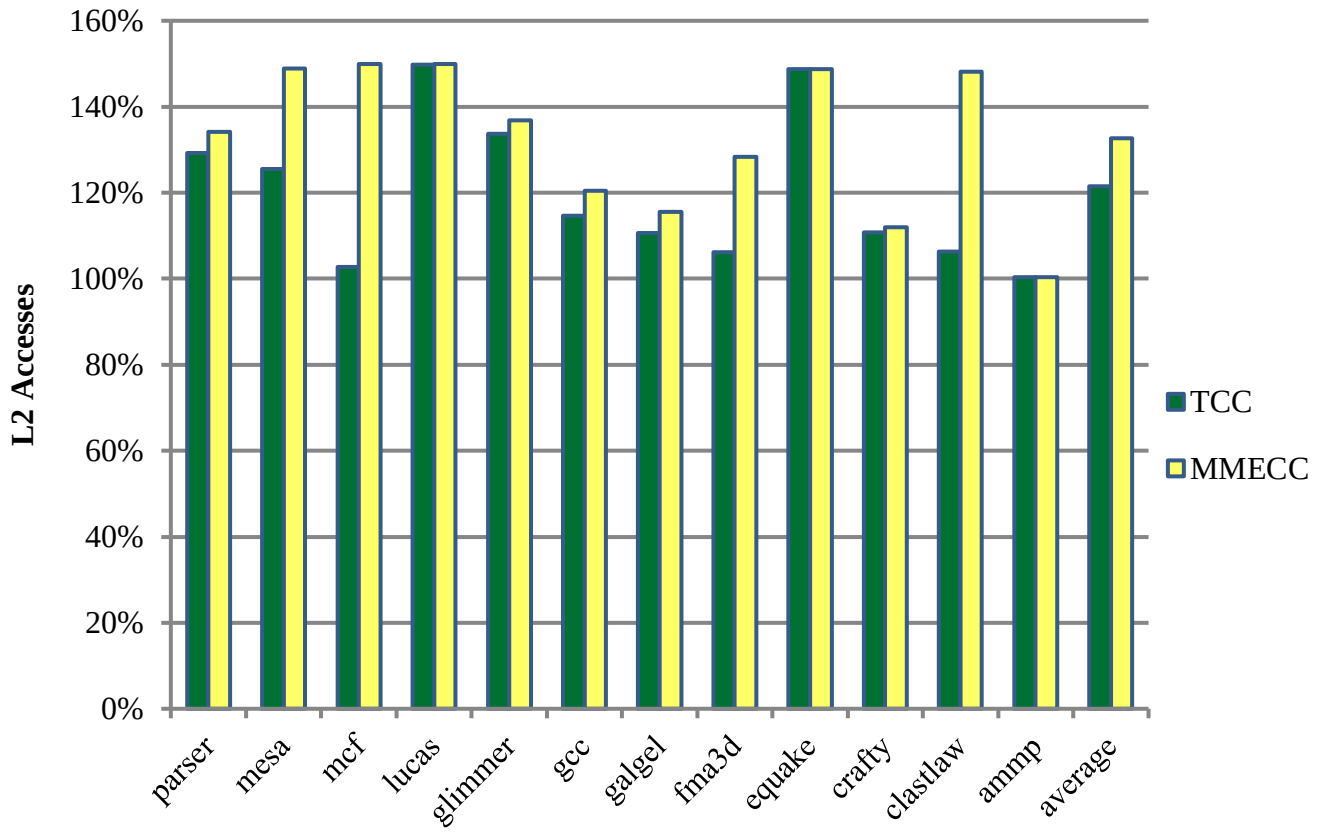

**Fig. 4.** L2 cache accesses frequency.

## 7.3 Energy

In this section we report energy consumption for L2 data cache for TCC and MMECC compared to the conventional system. We report dynamic and total energy in two following sections respectively.

**Dynamic Energy.** In **Fig. 5** we report L2 dynamic energy consumption for TCC and MMECE compared to the conventional system. MMECC shows an average increase of 8% (max: 22%) in energy consumption. Meantime, TCC shows an average reduction of 1% (max: 19%) and 9% (max: 32%) compared to the conventional and MMECC respectively. This is explained as follows. As shown in Section 7.2 MMECC shows higher cache access compared to the conventional system increasing cache dynamic energy. For TCC, L2 access frequency is higher than the conventional method. However, the conventional method stores ECC at the end of each cache block increasing cache energy consumption per access. On the other hand, TCC's cache access frequency is lower than MMECC. Therefore, TCC consumes less dynamic energy compared to MMECC.

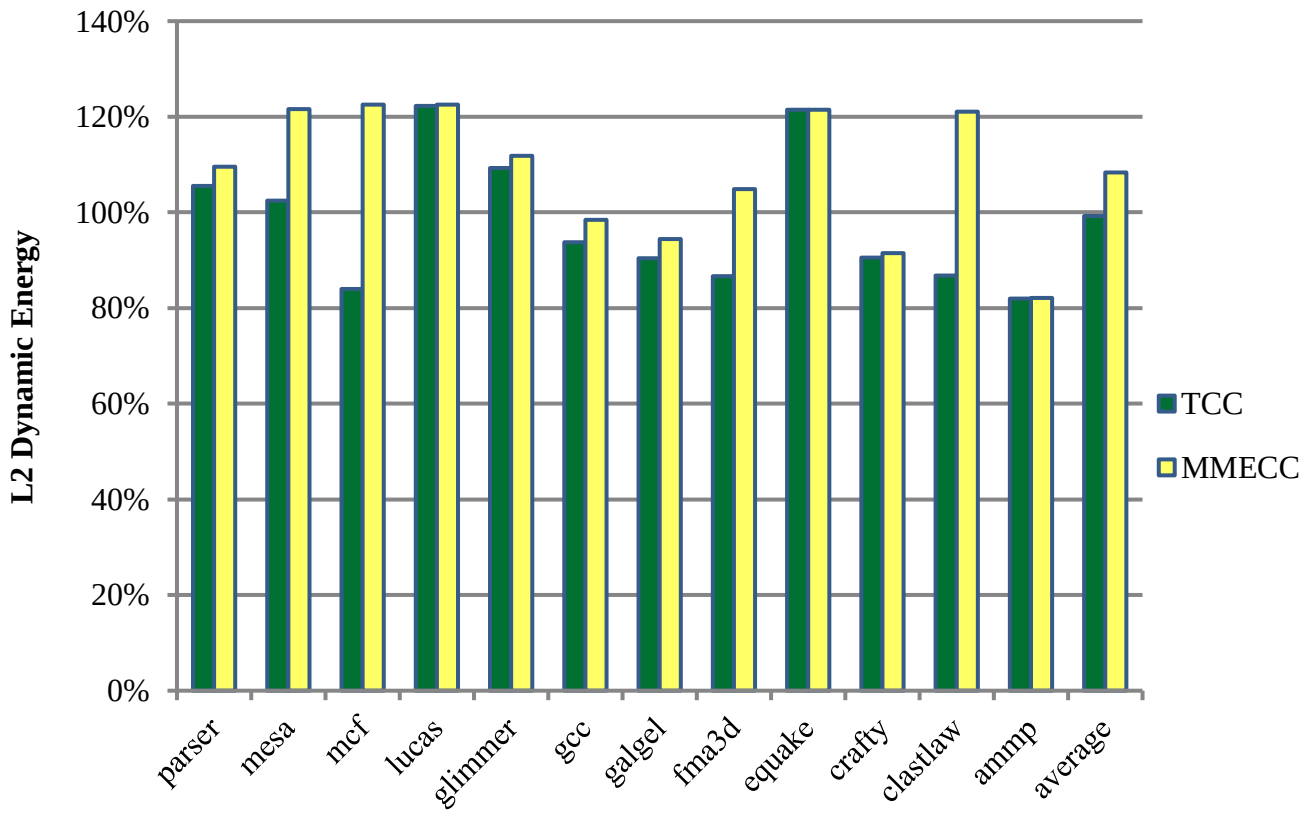

**Fig. 5.** Relative L2 data cache dynamic energy.

**Total Energy.** In **Fig. 6** we show L2 data cache total energy consumption for TCC and MMECC compared to the conventional system. We measure total energy as the summation of leakage and dynamic energy. As TCC and MMECC do not use a dedicated cache space to store ECC, they show 13% less leakage energy consumption compared to the conventional system. Meantime, TCC shows only 0.06% higher leakage energy compared to MMECC as TCC uses extra space to store signatures. Since TCC has low dynamic energy consumption (see Dynamic Energy Section), it shows 1% (up to 8%) and 12% (up to 13%) reduction in L2 data cache total energy compared to MMECC and conventional system respectively.

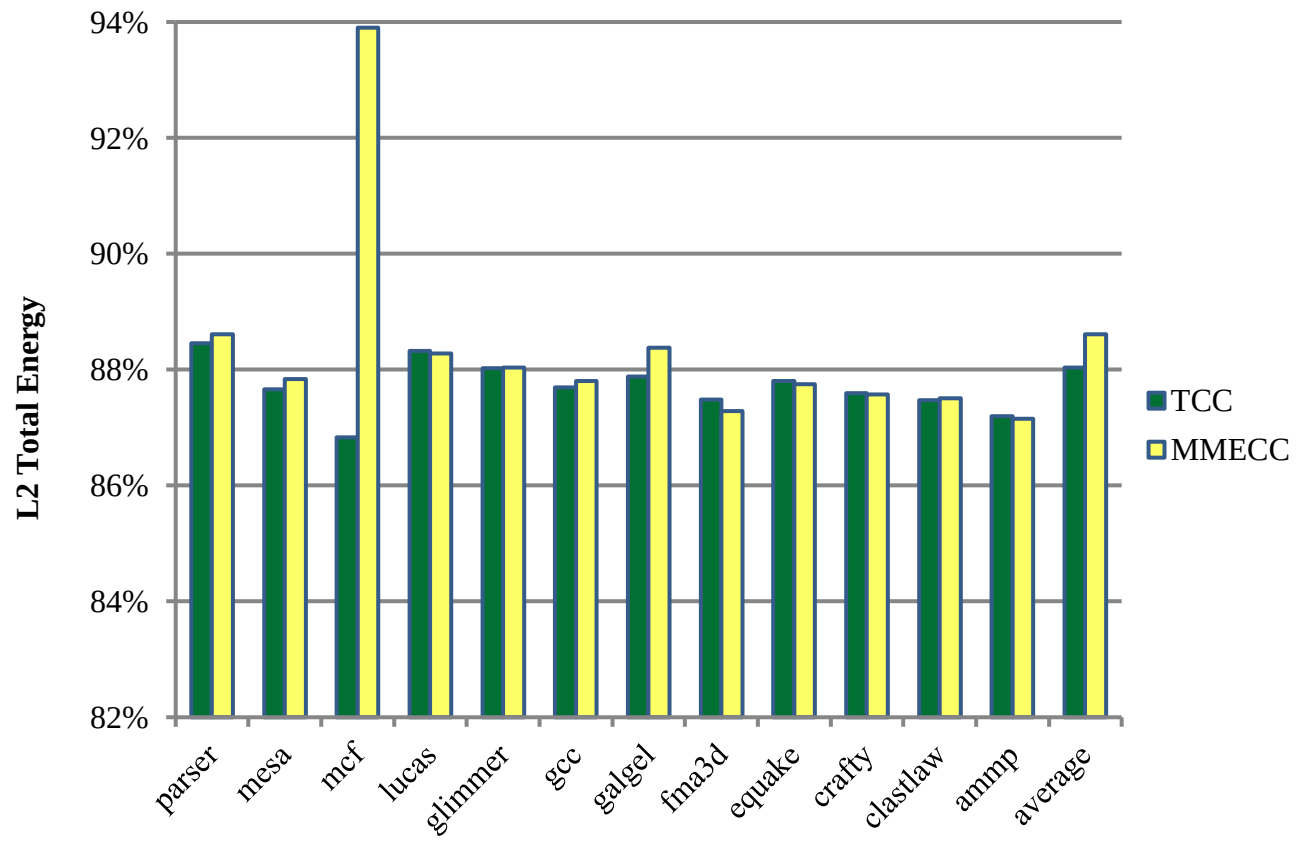

**Fig. 6.** Relative L2 data cache total energy consumption

### 7.4 Miss Rate

In **Fig. 7** we report L2 data cache miss rate for TCC and MMECC compared to the conventional system. TCC and MMECC show higher miss rate compared to the conventional method as they use some cache space to store ECC rather than using a dedicated cache. TCC shows 0.15% (up to 3%) less cache miss rate compared to MMECC. This reduction could be explained as follows. Both TCC and MMECC store ECC only for dirty cache lines. This results in an increase in the cache space occupied by ECC as the number of dirty cache lines increases. Meantime, TCC treats silent writes as clean writes as these writes do not change the value of the cache line. As TCC does not save ECC for these cache lines, ECC occupies less cache space for TCC compared to MMECC. TCC's miss rate is 0.7% (up to 3%) higher than the conventional system. MMECC's miss rate is 0.85% (up to 6%) more than conventional system. Note that the conventional system uses a dedicated cache array to store ECC.

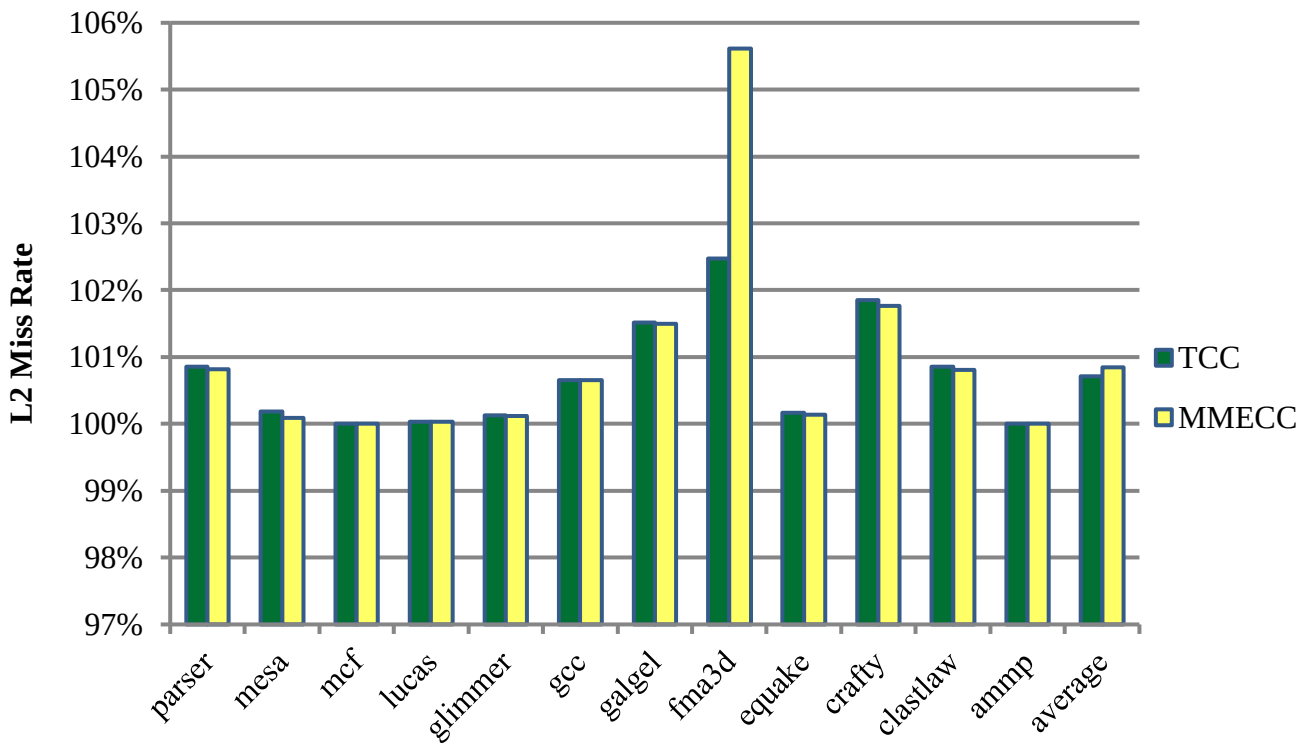

**Fig. 7.** L2 data cache miss rate

## 7.5 L1 Cache Size and Performance

In **Fig. 8** we show how variations in L1 data cache size impacts performance for TCC and MMECC compared to the conventional system. TCC shows better performance compared to MMECC for both 32KB and 128KB L1 data cache sizes.

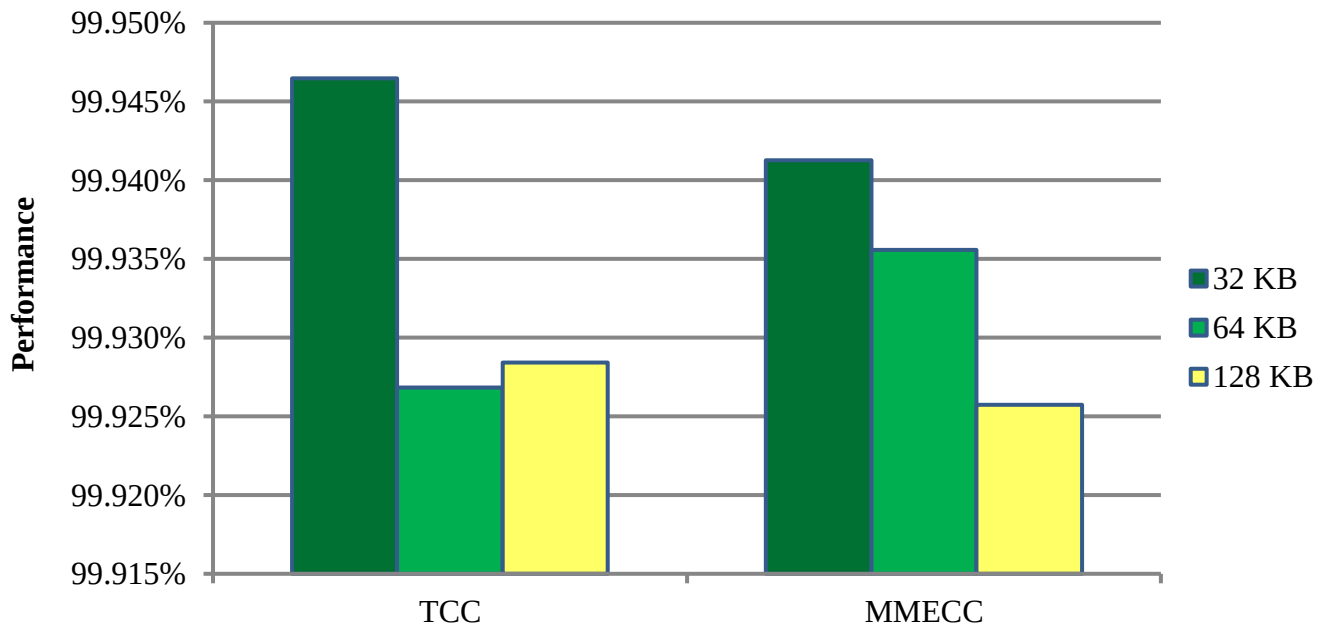

**Fig. 8.** Performance sensitivity to L1 data cache size.

## 7.6 L1 Cache Size and Total Energy

In **Fig. 9** we report how variations in L1 data cache size impacts total energy for TCC and MMECC compared to the conventional system. For all L1 data cache sizes, TCC shows lower total energy compared to both MMECC and conventional system.

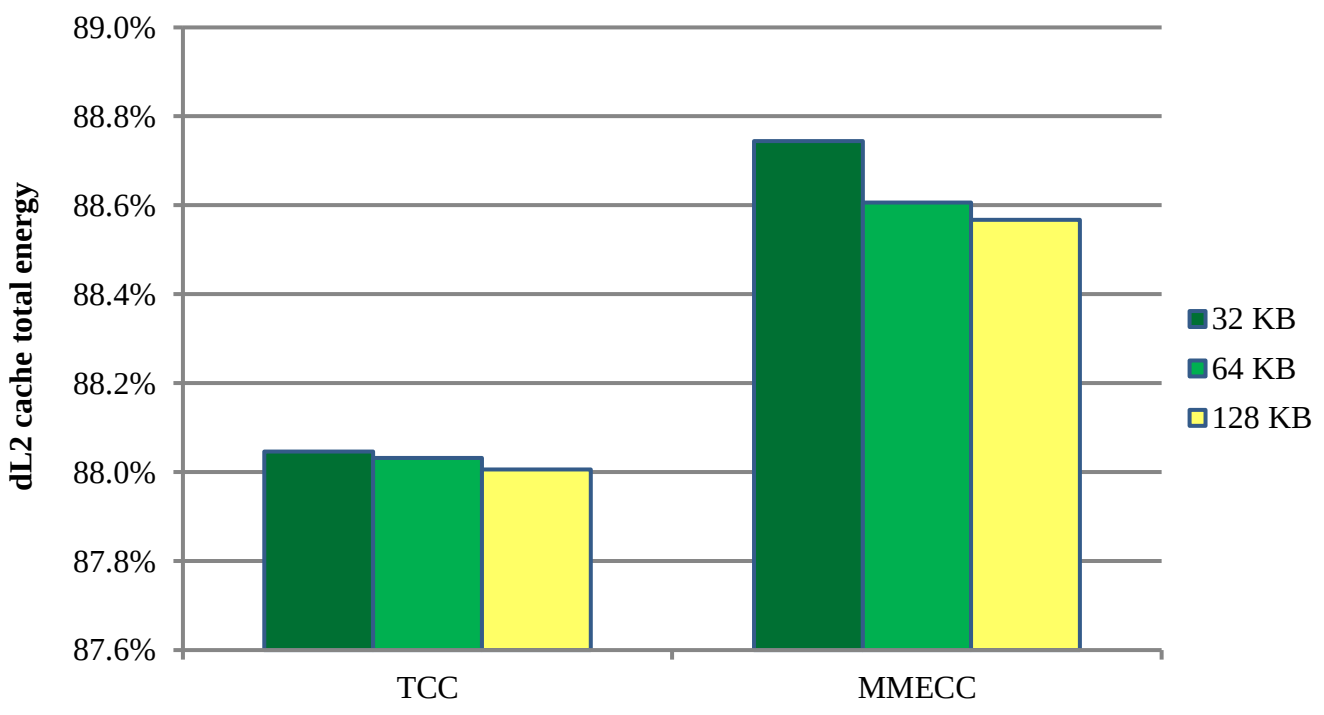

**Fig. 9.** Total energy sensitivity to L1 data cache size.

## 8 Conclusion

We showed that a significant portion of cache writes are silent, i.e., they write the data already existing in the cache. Based on this observation, we proposed Low-Power Traffic-Aware ECC to reduce cache traffic and access frequency. To detect silent writes, we use parity bits as signature bits and propose an efficient detection method. TCC avoids cache block write as well as ECC calculation and ECC write for silent writes. We consider the overhead of silent write detection in our experimental results and show that by taking silent writes into account, we improve both cache traffic and energy while maintaining performance.

## References

[1] Doe Hyun Yoon, Mattan Erez, "Memory mapped ECC: low-cost error protection for last level caches," International Symposium on Computer Architecture (ISCA), pp. 116-127, 2009.

[2] W. Zhang, S. Gurumurthi, M. Kandemir, A. Sivasubramaniam, "ICR: In-Cache Replication for Enhancing Data Cache Reliability," International Conference on Dependable Systems and Networks (DSN), June 2003.

[3] S. Kim and A. K. Somani, "Area Efficient Architectures for Information Integrity in Cache Memories," International Symposium on Computer Architecture (ISCA), May 1999.

[4] N. N. Sadler and D. J. Sorin, "Choosing an Error Protection Scheme for a Microprocessor's L1 Data Cache," International Conference on Computer Design (ICCD), October 2006.

[5] Lin. Li, V. S. Degalahal, N. Vijaykrishnan, M. Kandemir, and M. J. Irwin, "Soft Error and Energy Consumption Interactions: A Data Cache Perspective," International Symposium on Low Power Electronics and Design (ISLPED), August 2004.

[6] S. Kim, "Area-Efficient Error Protection for Caches," Conference on Design Automation and Test in Europe (DATE), March 2006.

[7] Kevin M. Lepak, Mikko H. Lipasti, "Silent stores for free," ACM/IEEE international symposium on Microarchitecture, pp. 22-31, December 2000.

[8] Doug Burger, Todd M. Austin, "The SimpleScalar tool set, version 2.0," ACM SIGARCH Computer Architecture News, vol. 25, no. 3, pp. 13-25, June 1997.

[9] [Online]. Available: http://www.spec.org/cpu2000/

[10] Naveen Muralimanohar, Rajeev Balasubramonian, Norman P. Jouppi, "Architecting Efficient Interconnects for Large Caches with CACTI 6.0," MICRO, vol. 28, no. 1, pp. 69-79, 2008.

[11] E. Perelman, G. Hamerly, M. V. Biesbrouck, T. Sherwood, and B. Calder, "Using SimPoint for Accurate and Efficient Simulation," ACM SIGMETRICS the International Conference on Measurement and Modeling of Computer Systems, June 2003.

[12] J. Karlsson, P. Ledan, P. Dahlgren, and R. Johansson, "Using heavy-ion radiation to validate fault handling mechanisms," MICRO, pp. 8–23, February 1994.

[13] J. Sosnowski, "Transient fault tolerance in digital systems," MICRO, pp. 24–35, February 1994.

[14] L.Z. Scheick, S.M. Guertin, G.M. Swift, "Analysis of Radiation Effects on Individual DRAM Cells," IEEE Transact. Nucl. Sci. vol. 47, No. 6, pp. 2534-2545, 2000.

[15] J. Standards. JESD89 Measurement and Reporting of Alpha Particles and Terrestrial Cosmic Ray-Induced Soft Errors in Semiconductor Devices, JESD89-1 System Soft Error Rate (SSER) Method and JESD89-2 Test Method for Alpha Source Accelerated Soft Error Rate, 2001.
[16] M. S. Gordon, K. P. Rodbell, D. F. Heidel, C. Cabral Jr., E. H. Cannon, D. D. Reinhardt, “Single-event-upset and alpha-particle emission rate measurement techniques,” IBM Journal of Research and Development, vol. 52, issue 3, May 2008.
[17] J. Maiz, S. Hareland, K. Zhang, and P. Armstrong, “Characterization of Multi-Bit Soft Error Events in Advanced SRAMs,” IEEE International Electron Devices Meeting (IEDM), December 2003.
[18] C. W. Slayman, “Cache and memory error detection, correction, and reduction techniques for terrestrial servers and workstations,” IEEE Trans. on Reliability, vol. 5, no. 3, pp. 397–404, Sept 2005.
[19] [Online]. Available: http://www.synopsys.com/tools/Verification/AMSVerification/CircuitSimulation/HSPICE/Pages/default.aspx
[20] R. Kalla, Sinharoy Balaram, J.M Tendler, “IBM Power5 chip: a dual-core multithreaded processor” MICRO, pp. 40-47, Mar-Apr 2004.
[21] [Online]. Available: http://www.amd.com/us/products/desktop/processors/athlon/Pages/AMD-athlon-processor-for-desktop.aspx
[22] [Online]. Available: http://software.intel.com/en-us/articles/recap-virtual-memory-and-cache/